\documentclass[11
pt,epsf,letterpaper]{article}%
\usepackage{amsmath}
\usepackage{geometry}

\usepackage{graphicx}
\begin{document}
\title{{\bf Generalized detailed balance relation for black hole horizon's fluctuations}}
\author{$^{\footnote{agomez@ucm.cl}}\ $Arturo J. Gomez and $^{\footnote{cpaiva@ucm.cl}} \  $Carlos Paiva $$\\\textit{Departamento de Matematica, Fisica y Estadistica, Facultad de Ciencias Basicas}, \\\textit{Universidad Catolica del Maule.}}
\maketitle

\begin{abstract}

In classical thermodynamics, irreversible processes are accomplished with an increase of entropy and a release of heat into the environment. In the case of black hole thermodynamics, instead, the increase of entropy is related with the absorption of energy from the outside of the black hole. Based on this analogy, the black hole is considered as a thermodynamic open system in the non-equilibrium regime during the absorption. Applying a novel result, we obtain a detailed balance relation for the macroscopic absorption of an astronomical object by an static spherically symmetric black hole. Because of the special characteristic of the thermodynamic system, it is necessary to include an additional term for the entropy produced by the absorption, allowing us to set a lower bound for the entropy production due to internal processes associated to the fluctuation. 
\end{abstract}

\section{Introduction} 

Since the seminal works  \cite{Bekenstein:1973ur}\cite{Bardeen: 1973gs}, it is well known that black holes after suitable identifications of the black hole parameters with usual thermodynamic variables obeys a kind of thermodynamic laws . One of the main results related to this theory is the computation of the black hole entropy, which turns out to be proportional to the area of the black hole horizon. Moreover, it was shown that, at classical level, any process involved in the increasing of the area of the horizon is an irreversible process, this fact gave the basis to formulate what is known as the Second Law of Black Hole Thermodynamics. Later on, Hawking showed that at semiclassical level, due to quantum effects at the horizon, the black hole behaves as a black body, releasing radiation in the thermal spectrum \cite{Hawking:1974rv}\cite{Hawking:1974sw}. In this way, it was shown that Hawking radiation implies that isolated black holes becomes thermodynamically unstable, affording even its evaporation. These results shows an apparent tension between General Relativity and Quantum Field Theory, known as information loss paradox \cite{Marolf:2017jkr}. If one think that information must be conserved, general relativity must be modified \cite{Almheiri:2012rt} and the internal microscopic degrees of freedom of the black hole, encoded by the quanta of the gravitational field, has to be considered \cite{Rovelli}\cite{Rovelli2}. 
Motivated by these facts, in this work we suggest that black holes should be considered as open systems, with a continuous exchange of matter and energy with the environment, characterized by the absorption of the surrounding matter and also, in a much smaller rate, by the effect of the Hawking radiation that contributes to the dissipation of the black hole.
\\ A fundamental feature of microscopic Hamiltonian dynamics is the time reversal symmetry. In the context of statistical mechanics of systems in contact with a thermal bath, a manifestation of this symmetry is the detailed balance relation at microscopic level. If we give label to micro-states of the system $i$,$j$, etc. and associate energies $E_{i},E_{j},etc.$, respectively with each such micro-state, the heat released into the environment over a transition of the system from $i$ to $j$ is given by $\Delta Q_{i\rightarrow j}=E_{i}-E_{j}$. Then, the quantity $\beta\Delta Q_{i\rightarrow j}$ is the variation of entropy of the bath over the course of the transition. Therefore, we may write \cite{Gardiner}, 

\begin{equation}
\frac{\pi(j\longrightarrow i)}{\pi(i\longrightarrow j)}=\exp{[-\Delta S_{i\rightarrow j}]}
\label{DBE}
\end{equation}

This microscopic expression relates irreversibility (in the sense of which process is more likely to happen) with the entropy produced by this process. Moreover, based on this hypothesis, several results has been established for systems into the non-equilibrium thermodynamic regime, such as the well known fluctuation theorem \cite{Gallavotti:1995zz}\cite{Crooks} among other important results as \cite{Jarzinsky:1996oqb}. %  \cite{Ruelle}\cite{England}.
%In the ase of the black hole, we shall consider the black hole horizon as the boundary of the system. The external sources of entropy or exchange channels of matter and energy with the environment are, in the more general case, the mass of the particles absorved its angular momenta, the  electric charge, among other process which could involve contributions to the variation of entropy
In this letter, based on a general result, analogous to Eq. (\ref{DBE}) in the context of to the black hole thermodynamics, we shall implement a recent scheme discussed in \cite{England} \cite{Ruelle2}\cite{Ruelle} to obtain a detailed balance relation for the absorption of a macroscopic astronomical object by a Schwarzchild black hole. 

In section II, we show a derivation of a detailed balance relation for macroscopic matter absorption based on results obtained from microscopic stochastic black hole horizon fluctuations. In section III we present a generalization of this result for deterministic dynamics by using the variation of the Gibbs free energy between the initial and final states of the transition. In section IV we present some physical consequences of this relation and finally, in section V we present our conclusions.

\section{Horizon fluctuations and macroscopic irreversibility}

The origin of the irreversibility described in the Hawking area law for the black hole's horizon, lies on the violation of the time-reversal symmetry imposed by the in-going boundary conditions of the black hole coordinates at the horizon. In a Schwarzchild space-time for semiclassical fields near to the horizon, the regularity of the Kruskal coordinates, requires that the out-going field modes must vanish, while the in-going fields doesn't.
On other hand, the microscopic transition probabilities between two different configurations of a Schwarzchild black hole horizon has been studied for the Hawking radiation process in \cite{Massar:1999wg}. Remarkably, the authors were able to show that the probability to emit a particle is governed by a change in the horizon area, given by $e^{(-\Delta A/4)}$, where $\Delta A/4$ is the decrease in the horizon area induced by the emission. In this work, as it was argued in \cite{Iso:2010tz} and \cite{Iso:2011gb}, we shall consider this relation also valid for the classical absorption process,

%Specifically, the ratio of the transition probability $\mathcal{P}$ between a configuration $\mathcal{C}$ with a black hole horizon area $A$ and another one $\mathcal{C^{'}}$ with corresponding area $A^{'}$ and the corresponding reversed process turns out to be governed by an amount in the horizon area,

\begin{equation}
\frac{\cal{\pi}(\cal{C^{'}(\cal{A^{'}})\rightarrow\cal{C(A)})}} {\cal{\pi}(\cal{C(A)}\rightarrow\cal{C^{'}(\cal{A^{'}}))}}=\exp{\left(\frac{\Delta \cal{A}}{4}\right)}
\label{BHDT}
\end{equation}
Where $\cal{C(A)}$ and $\cal{C^{'}(A^{'})}$ are two microscopic configurations of the horizon with different areas $\cal{A}$ and $\cal{A^{'}}$ and  $\Delta \mathcal{A}$ = $\cal{A^{'}}-\cal{A}$.

By recognizing that $\Delta \cal{A}$ is related to the energy $\Delta E$ by the First Law of black hole thermodynamics $(\Delta \mathcal{A})/4=(\Delta E)/T_{H}$, where $\Delta E$ is energy absorbed by the black hole from the environment in terms of mass, It is interesting to observe that the expression (\ref{BHDT}) is quite similar to the usual thermodynamic detailed balance, eq. (\ref{DBE}), with different sign in the flow of energy.
%Thus, (\ref{BHDT}) relates irreversibility (in the sense of which process is more likely) with the entropy produced by this process.
Also, based on Eq. (\ref{BHDT}), the authors \cite{Iso:2010tz} managed to provide a non-equilibrium fluctuation theorem, in the Crooks version, for classical matter absorption by a black hole. Moreover, based on this tool and taking account stochastic dynamics, the time evolution of the probability distribution function was described by the Fokker-Planck equation and the ratio is evaluated in a general Langevin processes. Using the methodology described in, they are able to obtain a generalized second law of black hole thermodynamics \cite{Iso:2011gb} 

In this work we are interested in the macroscopic consequences of equation (\ref{BHDT}), therefore, following \cite{England} we shall define more precisely, in terms of probabilities, the phenomena in which we are interested.
In $t=0$, the system is originally in the state $A$ defined by the macroscopic coarse-graining condition: \textit{The black hole has a mass $M$ and the corresponding area of the event horizon is $\cal{A}$}. At this point it is possible to associate a probability density in which the system is in a particular microscopic state $i$ within the macro-state $A$, denoted by $p(i/A)$.

Then, for $t=\tau\geq 0$, a macroscopic, uncharged massive object of mass $m$, like a massive planet or a neutron star, begins to be absorbed by the black hole and the system is driven out to a non-equilibrium regime over a time $\tau$ . After this fluctuation, the system reaches to a new steady state $B$, defined by the observable condition: \textit{The black hole has a total mass given by $M+m$ and the corresponding area of the event horizon has increased by a variation $\Delta \cal{A}$}'. Similarly, we can associate a probability distribution $p(j/B)$, as the likelihood of the system being in the micro-state $j$ given that originally the system was in $A$  and subsequently evolved over a time $\tau$ and it was then found in the macro-state $B$. It's important to remark that ($A,B$) are meta-stables states, whose dynamics is restricted to an isolated sector of the phase space described by equilibrium thermodynamics. Nevertheless, once the fluctuation takes place, the system is driven out through a non-equilibrium thermodynamic transition that connects the macro-states described above.

As we are interested in macroscopic irreversibility, which is encoded by the ratio between $\pi(A\longrightarrow B)$ and $\pi(B\longrightarrow A)$, we define these probabilities in terms of the microscopic variables

\begin{eqnarray}
\pi(A\longrightarrow B)=\int_{B}dj \int_{A}p(i/{A})\pi(i\longrightarrow j)di. 
 \\ 
\pi(B\longrightarrow A)= \int_{A}di \int_{B}p(j/B)\pi(j\longrightarrow i)dj.
\end{eqnarray}

Then, we define the configurations $\cal{C(A)}$ and  $\cal{C^{'}(A^{'})}$ as being the states  $i$ and $j$, respectively. By recognizing that the change of energy $\Delta E$ is the mass absorbed by the black hole $\Delta M_{i\longrightarrow j}$,
we take the ratio $R$ of these transition probabilities between macro-states \cite{England},
\begin{eqnarray}
\frac{\pi(B\longrightarrow A)}{\pi(A\longrightarrow B)}&=&\frac{\int_{A}di\int_{B}\left(\frac{p(j/B)}{p(i/A)}\right)p(i/A)\pi\left(j\longrightarrow i\right)dj}{\int_{B}dj\int_{A}p(i/{A})\pi(i\longrightarrow j)di}
\\
&=&\frac{\int_{A}di\int_{B}p(i/A)\pi\left(i\longrightarrow j\right)e^{\ln\left(\frac{p(j/B)}{p(i/A)}\right)}e^{\beta\Delta M_{i\rightarrow j}}dj}{\int_{B}dj\int_{A}p(i/{A})\pi(i\longrightarrow j)di}
\\
%e^{\ln\left(\frac{\pi(B\longrightarrow A)}{\pi(A\longrightarrow B)}\right)}
&=&\left<e^{\left(\beta\Delta M_{i\rightarrow j}+\ln\left[\frac{p(j/A)}{p(i/B)}\right]\right)}\right>_{A\longrightarrow B} 
\\
\left<e^{\left(-\ln{\frac{\pi(B\longrightarrow A)}{\pi(A\longrightarrow B)}}+ \ln{\frac{p(j/B)}{p(i/A)}}-\beta\Delta M_{i\rightarrow j}\right)}\right>
&=&1.
\label{ave}
\end{eqnarray}
%our aim is to perform the ratio between the forward and reversed transition probabilities, , respectively.
Here, $\left<...\right>_{A\longrightarrow B}$ represent the average over all paths from some $i$ in the initial ensemble $A$ to some $j$ in the final ensemble $B$, with each path weighted by its likelihood. Then, one can associate the Shannon entropy definition to the micro-ensembles as $S=\sum_{i}p_{i}\ln p_{i}$ that allow us to write
\begin{equation}
\ln\left(\frac{\pi(B\longrightarrow A)}{\pi(A\longrightarrow B)}\right)+\Delta S_{micro}-\beta\left<\Delta M\right>_{A\rightarrow B}\geq 0.
\label{MDB}
\end{equation}
%This relation is a macroscopic detailed balance relation between irreversibility, the microscopic entropy production associated  and the mass (energy) absorbed during the process.
The relation above is quite analogous to the one founded in \cite{England}, with the absorbed mass $\left<\Delta M\right>_{A\rightarrow B}=m$ playing the role of the heat released by the thermodynamical system during an irreversible process, but with the opposite flow.

Contributing to the discussion, in \cite{Ruelle} is pointed out that replacing the stochastic dynamics by deterministic dynamics and making use of meta-stable states in the phase space, the relation (\ref{MDB}) turns out to be an equality, giving an exact relation between the ratio $R$, the entropy produced by the microscopic components of the system and the mass exchanged with the environment. We shall discuss more this issue in the next sections.
%The left hand side of (\ref{ave}) is introduced trivially in the average, obtaining %$$
%A generalisation of this of relations, Eqs. (\ref{DB}) and (\ref{BHDT}), for macroscopic coarse-grained states connected by an irreversible transformation, is recently proposed in \cite{England},\cite{Ruelle},\cite{Relle2} in the context of biological systems. 

%Inspired in these results, we attempt to establish a generalised detailed balance relation for the process of classical matter absorption by the black hole. %The main idea is to describe the ratio of the transition probabilities for this phenomena by evaluating the variation of the Gibbs free energy between the starting and ending states of the transition, due to internal and external contributions.
%Also, we shall assume that the starting and ending states states corresponds to macroscopic coarse-graining observable conditions described by equilibrium thermodynamics, while during the transition time between these states, the system is driven out of the thermodynamical equilibrium.
%  

\section{General detailed balance for black hole absorption}

%A similar result was obtained in a different way by \cite{Ruelle}, 

In this section, we shall provide a generalization of the equation (\ref{MDB}) starting from the hypothesis that at equilibrium the black hole thermodynamics is governed by the
Gibbs free energy, (see \cite{Altamirano:2014tva} for a recent application) and that the ratio $\frac{\pi(B\longrightarrow A)}{\pi(A\longrightarrow A)}$ is proportional to the difference of the Gibbs free energy of the initial versus the final state. Also, following \cite{Ruelle}\cite{Ruelle2} it is assumed the microscopic time reversal symmetry and that the states $(A,B)$ are meta-stable states whose dynamics is constrained to an isolated region in the phase space.  %we shall obtain a similar result to (\ref{MDB})and of the transition generalizing the previous results \cite{England}\cite{Ruelle}.
. %So, here we shall make use of the previous results \cite{England}\cite{Ruelle}\cite{Ruelle2} to generalise the microscopic relation (\ref{DB}) to the coarse-grained macrostates (, $B$),
%between the macroscopic observable states $\cal{I}$ and $\cal{J}$, we shall obtain a detailed balance relation by generalising the ratio between forward and reverse transition probabilities given by

\begin{equation}
\frac{\pi(B\longrightarrow A)}{\pi(A\longrightarrow A)}=\exp{[\beta(G(B)-G(A))]}
\label{first}
\end{equation}
%Where $\mathcal{G}$ is identified with the Gibbs free energy of the black hole configuration. 
In general, for open systems, the total entropy variation can be written by \cite{Kondepudi}

\begin{equation}
dS=dS_i+dS_e
\end{equation}
where $dS_i$ is the internal entropy production and $dS_e$ is the entropy produced by the exchange of matter and energy through the action of an active environment. In our model, the black hole horizon is the boundary of the system and the interaction with the thermal bath is characterized by the matter that can be absorbed by the black hole, which we identify as an  uncharged mass $m$. We will assume that we can split the difference of the free energies into an internal part and another external due to the environment, which is characterized by some kind of microscopic number of channels labeled by $\alpha$. Defining the quantities $\pi^{\alpha}(A\longrightarrow B)$ as the specific probabilities of the channels $\alpha$ involved in the transition $A\longrightarrow B$ that satisfies the condition 

%charge, angular momenta, etc. of the particles absorbed by the black hole. This sources will be called external  $\emph{channels}$ following the construction \cite{Ruelle2}. Also, we assume that the transition probability $\pi(\cal{I}\longrightarrow \cal{J})$ is the sum of the probabilities of the channels, $\alpha$, of interaction with the environment
%\begin{equation}
%\pi(\mathcal{I}\longrightarrow \mathcal{J})=\sum_{\alpha}{\pi^{\alpha}(\cal{I}\longrightarrow \cal{J})}
%\label{alpha}
%\end{equation}

\begin{equation}
\frac{\pi^{-\alpha}(B\longrightarrow A)}{\pi^{\alpha}(A\longrightarrow B)}=\exp{[\beta\left(\Delta G^{int}+\Delta G^{\alpha}\right)]},
\end{equation}
where $\sum_{\alpha}\pi^{\alpha}(A\longrightarrow B)= \pi(A\longrightarrow B)$, it can be shown \cite{Ruelle} that the expression (\ref{first}) generalizes to
\begin{equation}
\frac{\pi(B\longrightarrow A)}{\pi(A\longrightarrow B)}=\sum_{\alpha}p^{\alpha}\exp{[\beta\left(\Delta G^{int}+\Delta G^{\alpha}\right)]}.
\label{second}
\end{equation}
%such that $\Delta G^{ext}\equiv \Delta G^{\alpha}$
%and
%\begin{equation}
%\pi(A\longrightarrow B)=\sum_{\alpha}\pi^{\alpha}(A\longrightarrow B)
%\label{alpha}
%\end{equation}
%The relation(\ref{first}) also is valid for the channels , which takes the form \cite{Ruelle}
%\begin{equation}
%\frac{\pi^{-\alpha}(B\longrightarrow A)}{\pi^{\alpha}(A\longrightarrow B)}=\exp{[\beta\left(\Delta G^{int}+\Delta G^{\alpha}\right)]}
%\end{equation}
%Note that if $\alpha=0$, we recover the expression (\ref{first}). The quantity $\Delta^{\alpha} G$ is the variation of the Gibbs free energy due to the absorption of energy by sources of mass, charge, angular momentum or other channels of energy exchange with the environment
In (\ref{second}) we have introduced the relative probabilities of $p^{\alpha}=\frac{\pi^{\alpha}(A\longrightarrow B)}{\pi(A\longrightarrow B)}$, enconding the information about the likelihood of the differents channels $\alpha$. Note that if $\alpha=0$, we recover the standard expression (\ref{first}).

%\begin{equation}
%p^{\alpha}(A\longrightarrow B)=\frac{\pi^{\alpha}(A\longrightarrow B)}{\pi(A\longrightarrow B)}.
%\label{p}
%\end{equation}

In black hole thermodynamics, the Gibbs free energy is given by $G=M -TS$, where the mass $M$ plays the role of the enthalpy. Thus, the ratio takes the form %and manipulating the relations (\ref{first}), (\ref{alpha}) and (\ref{second}), one can find a detailed equation for the ratio
\begin{equation}
\frac{\pi(B\longrightarrow A)}{\pi(A\longrightarrow B)}=\exp[-\Delta S_{int}]%\sum_{\alpha}p^{\alpha} 
\sum_{\alpha}p^{\alpha}\exp[\beta\Delta M^{\alpha}-\Delta S^{\alpha}]. 
\label{DB0}
\end{equation}

In this expression, $T$ is the black hole temperature. Also we have assumed that the black hole doesn't produce mass due to internal process, $i. e.$, $\Delta M_{int}=0$, instead we allow the possibility of internal entropy production $\Delta S_{int}$. %As we shall see, the inclusion of this term turns out to be very relevant to compensate because of the %besides of the usual black hole entropy proportional to the horizon area.

Then, recalling the convexity of the exponential function, we can write 

\begin{eqnarray}
\frac{\pi(B\longrightarrow A)}{\pi(A\longrightarrow B)}&\geq&\exp[-\Delta S_{int}+\sum_{\alpha}p^{\alpha} 
(\beta\Delta^{\alpha}M-\Delta^{\alpha} S)]. \\
&=&-\left<\Delta S\right>+\beta\left<\Delta M\right>,
\label{DetailedBH}
\end{eqnarray}

%By using the convexity of the exponential function we find

%\begin{equation}
%\frac{\pi(\cal{J}\longrightarrow \cal{I})}{\pi(\cal{I}\longrightarrow \cal{J})}\geq\exp[-T\Delta S_{int}+\sum_{\alpha}p^{\alpha}\left(\Delta^{\alpha}M+T\Delta^{\alpha}S\right)].
%\label{DetailedBH2} 
%\end{equation}
where we have defined $\left<\Delta S\right>= \Delta S^{int}+\sum_{\alpha}\Delta^{\alpha} S$ and $\left<\Delta M\right>=\sum_{\alpha}p^{\alpha}M^{\alpha}$. 

Remarkably, Eq. (\ref{DetailedBH}) is quite similar to the relation (\ref{MDB}), obtained in the previous section by a rather different way. 
Both equations corresponds to macroscopic detailed balance relations for the process of absorption of matter by the black hole, which are analogous to the ones proposed in $\cite{Ruelle}$ and $\cite{England}$ for the irreversible process of biological self-replication. %In particular, Eq. (\ref{DetailedBH}) is a particular case of a general relation proposed in \cite{ruelle2}.

In the specific case of a Schwarzchild black hole, for simplicity we can assume that the channels $\alpha$ has the same probability $p^{\alpha}$ to absorb the exterior mass $m$. This allows to factorize the sum over the probabilities $p^{\alpha}$  obtaining $\sum_{\alpha}p^{\alpha}=1$. Thus, from Eq. (\ref{DB0}) we have 

\begin{eqnarray}
\frac{\pi(B\longrightarrow A)}{\pi(A\longrightarrow B)}&=&\exp[-\Delta S_{int}]%\sum_{\alpha}p^{\alpha} 
\exp[\beta\sum_{\alpha}(\Delta M^{\alpha}-\Delta S^{\alpha})]. \\
&=&%\sum_{\alpha}p^{\alpha} 
\exp[-\Delta S_{int}-\Delta S_{hor}+\beta m]
\label{DB0}
\end{eqnarray}

It's important to remark that $S_{int}$ and $S_{hor}$ are associated to microscopic processes within the black hole and in the horizon, respectively. In this sense, the ratio $\frac{\pi B\longrightarrow A)}{\pi(A\longrightarrow B)}$ measures the irreversibility of a macroscopic event in terms of the microscopic entropy production and the variation of mass associated to the transition process. As we shall see, the inclusion of $-\Delta S_{int}$ turns out to be very relevant because of the presence of the term $\beta m$, which is interpreted as the energy absorbed by the black hole as a thermodynamic system.
Remarkably, (\ref{DB0}) provide us an exact detailed balance relation, which should not be necessarily valid for other black holes, as rotating or charged ones, once the relative probabilities of the absorption channels are in the rate  $m:e:\omega$ \cite{Hawking:1974sw}.
It's worthy to observe that both equations (\ref{MDB}) and (\ref{DetailedBH}) are very similar in their content, nevertheless its origin is quite different, in (\ref{MDB}) the mass is interpreted as the absorbed energy from the environment, analogous to the heat released by thermodynamic irreversible processes. Instead, in (\ref{DetailedBH}) the variation of mass is related to the variation of enthalpy of the system, encoded by the difference of the Gibbs free energy.

\section{The bound for the internal entropy production}
The applicability of the relations Eqs. (\ref{MDB}) and (\ref{DB0}), depends directly on the estimation of the time reversed probability $\pi(B\longrightarrow A)$.
It is important to remark that the irreversibility of the transition implies that the dynamics of the time-reversed process, $B\longrightarrow A$, in the phase space becomes highly unstable, so that the likelihood $\pi(B\longrightarrow A)$ turns out to be very difficult to determine. However, to solve this problem we shall use of a physically possible mechanism to give a bound for this reversed transition probability. To ilustrate this, let's suppose that $\pi(B\longrightarrow A)^{*}$ is a physically possible bound for the time-reversed transition probability. If we assume that all the entropy of the system lies in the horizon according to the holographic principle paradigm, equation (\ref{DB0}) takes the form

\begin{equation}
\ln\left(\frac{\pi(B\longrightarrow A)^{*}}{\pi(A\longrightarrow B)}\right)\geq-\frac{\Delta A}{4}+\frac{m}{T_{BH}}
\label{2lawBH}
\end{equation}

Naturally, it is assumed that $\pi(B\longrightarrow A)^{*}$ is several orders of magnitude smaller than $\pi(A\longrightarrow B)$, being the signature of the level of irreversibility of the process. 

As it is apparent from the last equation, if the temperature of the black hole $T_{BH}$ is low enough, the left hand side of Eq. (\ref{2lawBH}) will be larger than the logarithm of the right hand side and the inequality will be broken. Therefore, it is necessary to include an additional entropy $S_{int},$ attributable to internal degrees of freedom of the black hole to compensate the large value of $\beta m$ and hold the inequality (\ref{2lawBH}),

\begin{equation}
\ln\left(\frac{\pi(B\longrightarrow A)^{*}}{\pi(A\longrightarrow B)}\right)\geq-\Delta S_{int}-\frac{\Delta A}{4}+\frac{m}{T_{BH}}
\label{2lawBH2}
\end{equation}
%Until now, we have not talk about the presence of another entropy production $\Delta S_{int}$, besides the usual entropy proportional to the variation of horizon area $\Delta S_{hor}=\frac{\Delta A}{4}$

%=\frac{A}{4}$, 

%\begin{equation}
%\ln\left(\frac{\pi(\cal{J}\longrightarrow \cal{I})}{\pi(\cal{I}\longrightarrow \cal{J})}\right)=-\Delta S_{int}-S_{hor}+\beta m
%\end{equation}

%At this point, we identify the entropy due to the absorption of mass with the variation of the black hole horizon's area, $S_{hor}=\frac{1}{4}\Delta A$. Also, taking account that the Hawking temperature is $T_{H}=\frac{\kappa}{2\pi}$ and recalling that for the Schwarzchild black hole, the surface gravity is $\kappa=\frac{1}{4M}$, we can obtain the following relation,
%\begin{equation}
%\ln\left(\frac{\pi(\cal{J}\longrightarrow \cal{I})}{\pi(\cal{I}\longrightarrow \cal{J})}\right)\geq-\frac{\Delta S_{int}}{8\pi M}+\frac{\Delta A}{16\pi M}+m
%\label{Detailed3}
%\end{equation}

%Certainly the forward transition probability $\pi(\cal{I}\longrightarrow\cal{J})$ is several orders of magnitude greater than the reversed one. 

In order to implement the inequality (\ref{2lawBH2}) we must to obtain a value for $\pi(B\rightarrow A)^{*}$.  %in classical black hole thermodynamics the area law states that any process that increase the horizon area is an irreversible process
The time-reversed transition probability corresponds to the black hole give off all the mass absorbed throughout the horizon, decreasing its area until the original configuration of the black hole and the environment, over the same period of time $\tau$. Thus, the only physically allowed mechanism to shrink the area of the horizon is the Hawking radiation, $i.e$ to consider the probability of evaporation of the black hole horizon due to the steady conversion of quantum vacuum fluctuations into pair of particles at the horizon, as a bound for the time-reversed transition probability $\pi(B\longrightarrow A)$. The evaporation time, is proportional to the cubic of the black hole mass ($t_{evap}\sim M^3$). So, we shall estimate bound of the time reversed transition probability, as being the inverse of the time that takes the black hole in the state $B$, to evaporate the absorbed mass $m$, %$\pi(B\rightarrow A)^{*}$

\begin{equation}
\pi(B\longrightarrow A)^{*}\simeq\frac{1}{\tau_{evap}}\sim \frac{1}{m^3}
\end{equation}
%Where $m$, is the total mass absorbed by the black hole.
Finally, we can use the relation (\ref{2lawBH2}) to set a bound for the entropy production for the absorption process due to internal unknown degrees of freedom of the black hole.
\begin{equation}
\Delta S_{int}\geq
-2\pi m^{2}+3\ln m+\frac{m}{T_{BH}}
\label{End}
\end{equation}

Let us note, that L.H.S. of equation (\ref{End}) depends on the behavior of the black hole temperature, which in the case of Schwarzchild is inversely proportional to the black hole mass. 
If we have a very cold black hole, which absorbs an astronomical object with a small mass compared with the mass the black hole, the bound of the internal entropy production is large and grows linearly with the black hole mass $M$.  On other hand, if the temperature of the black hole begins to increase, the bound on $\Delta S_{int}$ become smaller.

\section{Conclusion}

In this work, we have considered a black hole interacting with a thermal bath as a thermodynamic open system. Based on Hamiltonian time reversal symmetry and on the microscopic detailed balance at the equilibrium, we have applied a recent result developed for thermodynamic irreversible phenomena\cite{England} \cite{Ruelle} to the process of classical matter absorption by a stationary and spherically symmetric black hole. 
We found a general detailed balance relation (\ref{DetailedBH}) between the degree of macroscopic irreversibility of the process, encoded by the logarithm of the ratio $R$ of the transition probabilities, the microscopic internal entropy associated to the transition and the energy absorbed in the process. In this relation, the absorbed mass plays the role of the enthalpy, which in the case of standard thermodynamics is released into the thermal bath along with the irreversible process.  As a consequence of the difference of sign in (\ref{End}), an additional term has to be considered for the entropy production, which is assumed to take account the entropy production due to internal degrees of freedom of the black hole. Remarkably, it turns out that this entropy increases linearly with the black hole mass, possibly achieving a large value for supermassive black holes absorbing astronomical objects with a much smaller mass $(M>>m)$. If we assume the argument that in non-equilibrium statistical physics, stable steady states may be maintained and self-organized under instabilities by the internal entropy production \cite{Nicolis, Borckmans}. One can speculate that a large value of $S_{int}$ should imply some kind of internal structure in the black hole, in agreement with the discussion in \cite{Almheiri:2012rt}. We think that this results also could be useful for the black hole information paradox problem.

As a future work, we can consider to generalize this result for black holes with a charge $Q$ and angular momentum $J$. In this case, all the relevant functions presents in Eq. (\ref{DetailedBH}), as well as the microscopic channels probability $p^{\alpha}$ shall depend on this additional black hole parameters.
 %The dependence of the black hole temperature, which is inversely proportional to the black hole mass, has consequences in the internal entropy production, because of sign of the R.H.S. of equation (\ref{2lawBH2})

\section*{Acknowledgments}

The authors would like to thanks Fabrizio Canfora, Adolfo Cisterna and Ingrid Carvacho for the critical reading and useful comments about the paper

\end{document}